\documentclass[aps,pre,superscriptaddress,nofootinbib,preprintnumbers]{revtex4-2}

\usepackage{graphicx}
\usepackage{graphics}
\usepackage{amsmath,amssymb}
\usepackage[usenames]{color}
\usepackage{subfigure}
\usepackage{hyperref}
\usepackage{enumitem}
\usepackage{lgreek}
\usepackage{float}

\newcommand{\be}{\begin{equation}}
\newcommand{\ee}{\end{equation}}
\newcommand{\bea}{\begin{eqnarray}}
\newcommand{\eea}{\end{eqnarray}}
\newcommand{\ben}{\begin{enumerate}}
\newcommand{\een}{\end{enumerate}}
\newcommand{\bit}{\begin{itemize}}
\newcommand{\eit}{\end{itemize}}

\newcommand{\la}[1]{\label{#1}}

\newcommand{\Fig}[1]{Fig.~\ref{#1}}

\newcommand{\vv}[1]{\mathbf #1}							

\newcommand{\bert}{\raise-0.45mm\hbox{\Large$\Box$}}			

\makeatletter
\newcommand*\bigcdot{\mathpalette\bigcdot@{.5}}
\newcommand*\bigcdot@[2]{\mathbin{\vcenter{\hbox{\scalebox{#2}{$\m@th#1\bullet$}}}}}
\makeatother

\definecolor{BrickRed}{cmyk}{0,0.89,0.94,0.28}					
\definecolor{MidnightBlue}{cmyk}{0.98,0.13,0,0.43}				
\definecolor{DarkGreen}{rgb}{0.100806,0.495968,0.209979}
\definecolor{orange}{rgb}{0.587167,0.354498,0.146197}

\graphicspath {{figures/}}

\begin{document}

\title{Thermal Vacuum Model for Cosmology without Inflaton  }
\author{Robert Alicki}
\email{robert.alicki@ug.edu.pl}
\affiliation{International Centre for Theory of Quantum Technologies (ICTQT), University of Gda\'nsk,  80-308, Gda\'nsk, Poland}

\date{\today}

\begin{abstract}
The previously proposed modification of the standard (flat) inflationary $\Lambda CDM$ model  in which the inflaton field(s) and ``dark energy"  are replaced by  the  vacum in expanding Friedmann-Lema\^itre-Robertson-Walker Universe is studied. The expanding joint vacuum of the all ingrediences of matter, including Standard Model particles and a dark matter sector, is treated as a thermal equilibrium state at temporal Gibbons-Hawking temperature, proportional to the Hubble parameter, and chemical potentials equal to particle masses. This theory applied to the early Universe provided  the new mechanism of inflation and its graceful exit accompanied by particles production.  Here, it is shown that the same idea applied to the late Universe explains  acceleration of expansion and gives a natural solution to the ``cosmological constant problem''.  Moreover, this formalism can be combined with the anomalous quantum gravity effects leading to  a viable baryogenesis mechanism  and  producing certain restrictions on the mass spectrum for dark matter particles. 

\end{abstract}

\maketitle

\tableofcontents

\newpage


\section{Introduction} 
\la{sec:intro}

In the previous papers  \cite{QTdS} , \cite{relax} a modification of the standard   flat inflationary   $\Lambda$CDM model was proposed, which instead of inflaton field(s),  cosmological constant  and reheating mechanism employs a single object - thermal energy of expanding vacuum. The idea that the vacuum of a quantum field in the expanding de Sitter Universe is related to the thermal state at the Gibbons-Hawking temperature 
\be
T_{\rm dS} = \frac{h}{2 \pi} .
\la{eq:dS}
\ee
proportional to the Hubble parameter $h$ has been introduced in \cite{Gibbons}. However, its physical interpretation was unclear  as well as its cosmological consequences.  A model describing a quantum system (e.g. atom or harmonic oscillator) interacting with a massless scalar field at the initial vacuum state in the de Sitter Universe has been studied in \cite{QTdS}. Applying standard methods of the quantum theory of open systems, including Markovian approximation in the weak coupling regime, it was shown that the vacuum acts on a quantum system (``thermometer'') as a thermal bath of  radiation.
\par
This idea can be extrapolated to massive fields with boson and fermion statistics and to interacting systems in the form of the following Thermal Vacuum (TV) hypothesis:

\emph{The energy density of the vacuum in the de Sitter space-time, measured in the comoving reference frame, is equal to the energy density of  all matter components at the thermal equilibrium state characterized by the Gibbons-Hawking temperature and the chemical potentials equal to the corresponding particle masses.}

Some remarks, concerning the above hypothesis, are in order :

\begin{enumerate}

	\item as shown by Sewell \cite{Sewell} the notion of thermal equilibrium is not covariant and makes sense only in the rest frame, what in this case means cosmic, comoving reference frame being at rest with respect to Cosmic Microwave Backgound (CMB),

	\item thermal energy density is well defined, in contrast to cut-off dependent  zero-point energy (which is renormalized out by normal ordering), and is computable in extreme cases of very high and very low temperatures using the methods of statistical mechanics for ideal quantum gases,
	
      \item putting chemical potentials equal to masses allows to keep $T_{\rm dS} $ as the only external (geometric) parameter characterizing the vacuum in the de Sitter space-time,

	\item one assumes that during the evolution  thermalization processes are fast enough to mantain thermal equilibrium of the vacuum with the time dependent temperature  $T_{\rm dS}(t) = h(t)/2\pi$,

      \item thermal energy of expanding vacuum plays the role of    ``inflaton's false vacuum energy'', ``dark energy'' or  `` cosmological constant''.

\end{enumerate}
	
In the following the proposal based on the hypothesis of above will be called Thermal Vacuum Model (TVM). Firstly, the results of \cite{relax} are reminded, where TVM was used to propose a  mechanism of inflation and its graceful exit without invoking inflaton and reheating processes. Then the case of late Universe will be studied using TVM in the limit of very low temperatures  to show that this model predicts observed acceleration of expansion and provides a certain bound on the maximal mass of dark matter particles. For the early universe  TVM can be combined with anomalous quantum gravity effects leading to  a viable baryogenesis mechanism. On the other hand,  for the late Universe, the similar mechanism allows to propose a consistent picture of dark matter sector with bounds on the particles  masses. 

\section{Friedmann equations for TVM}
\la{sec:relax}
The cosmology of the flat, uniformly isotropic and homogeneous Universe is described by  the Friedmann-Lema\^itre-Robertson-Walker (FLRW) metric
\be
ds^2 = - dt^2 + a^2 (t) \, d \vv x^2.
\la{eq:metric}
\ee
with the scale factor $a(t)$ normalized to one at present time and $t$ being the cosmological time. The evolution equations for the FLRW metric are naturally formulated in terms of the Hubble parameter
\be
h(t) = \frac{\dot a(t)}{a(t)}.
\la{eq:hubble}
\ee
Namely,  Einstein equations of General Relativity applied to FLRW metric are reduced to a pair of Friedmann equations expressed in \emph{Planck units} in the following way
\be
\frac{3}{8\pi}h^2 = \rho , \quad \dot h = -\frac{3}{2} h^2 - 4\pi p .
\la{eq:Fried}
\ee
Here, $\rho$ is the energy density and $p$ is the pressure of  the``cosmic fluid".
\par
The only practical difference between TVM and standard  inflationary cosmological models (see eg.\cite{Inflation} for a collection of about 120 ``simplest''  models)  are  the proposed formulas for  $\rho$  and $p$
\be 
\rho = \rho_m + \rho_{\rm dS},\quad p = p_m + p_{\rm dS}  ,
\la{eq:epTV}
\ee
where  $\rho_m$ is the energy density of  matter (including regular and dark matter) and  $\rho_{\rm dS}$ is the energy density of TV,  while $ p_m$ and $p_{\rm dS} $ are corresponding pressures, respectively. 
In the next step one needs to establish equations of state which connect energy densities and pressures. The relation for matter is usually written in the form
\be
 p_m  = w_m \rho_m  ,
\la{eq:parw}
\ee 
where $w_m$ is, generally, a complicated function dependent on the temporal composition of  matter fluid and corresponding temperatures of different components. In the two simplest cases of highly relativistic matter (``radiation'') \hbox{$w_m \simeq 1/3$} while for nonrelativistic one  \hbox{$ w_m \simeq 0 $}. The equation of state for TV component reads
\be
p_{\rm dS} = -\rho_{\rm dS} ,
\la{eq:pressuredS}
\ee
and expresses the fact that the TV energy density is not diluted by  space expansion.  This property is shared with the cosmological constant, inflaton energy density in the slow-roll regime 
or generally ``dark energy''.
\par
Combining \eqref{eq:Fried}  -  \eqref{eq:pressuredS}  one obtains the basic equations of the TVM
\be
\rho_m + \rho_{\rm dS}  - \frac{3}{8\pi }h^2 = 0 ,
\la{eq:1FE}
\ee
\be
\dot h =  - 4\pi (1 + w_m)\,\rho_m  .
\la{eq:dh}
\ee
The eq.\eqref{eq:1FE} can be interpreted as the total energy density balance where a sum of positive matter energy density and positive TV energy density  is completely compensated by the negative contribution from gravity of the curved FLRW space-time, while  \eqref{eq:dh} is a dynamical equation for the Hubble parameter. Notice, that in contrast to the standatd $\Lambda$CDM model the vacuum energy density $\rho_{\rm dS}$ is not a constant, but is a function of $h(t)$. Although $\rho_{\rm dS}$ is not diluted by expansion, it  interacts with the matter density $\rho_m$, in particular  massively creating regular and dark matter at the end of inflation period (see the next subsection). It follows from \eqref{eq:1FE} and \eqref{eq:dh} that positivity of $\rho_m$ and the restriction $w_m > -1$ imply that the Hubble parameter is a monotonically decreasing function of time.

\subsection{Early evolution of the Universe}
\la{sec:early}
At the very early stage of the evolution one expects that $T_{\rm dS}$ is much higher than any elementary particle mass including dark matter components. Thus TV can be approximately treated as radiation
with Stefan-Boltzmann energy density proportional to $T^4_{\rm dS}$ such that
\be
\rho_{\rm dS} = \sigma h^4, \quad \sigma = \frac{g_f}{480 \pi^2} .
\la{eq:rhodS1}
\ee
Here $g_f$ is the total number of ``polarizations" with fermion contribution weighted by a factor $7/8$ \cite{K&T}. Under the hypothesis that the energy exchange between baryonic matter and dark matter is negligible during the evolution up to the present time, and taking the estimated present ratio of dark matter energy density to baryonic one as about 5 one gets  $\sigma\simeq 0.15$. Indeed, while the Standard Model of elementary particles provides about 100 polarizations, one should add about 500 polarizations for a dark matter particles to obtain $g_f\simeq 600$.
\par
As long as the Gibbons-Hawking temperature is well above the highest mass scale we can use simplified Friedmann equations putting in \eqref{eq:1FE}, \eqref{eq:dh} the formula \eqref{eq:rhodS1}  and $w_m = 1/3$,  obtaining
\be
\rho_m  = \frac{3}{8\pi}h^2 \left[1 - \left(\frac{h}{h_{0}}\right)^2\right] ,\quad \dot h =- \frac{16\pi}{3} \rho_m ,
\la{eq:1FE1}
\ee
and
\be
\rho_{\rm dS} = \frac{3}{8\pi} \frac{h^4}{h_0^2},\quad \rm{with}\quad  h_0 \equiv \sqrt{\frac{3}{8\pi \sigma}}\simeq 0.9 .
\la{eq:rhodS2}
\ee
Equations \eqref{eq:1FE1} possess two stationary solutions  with $\rho_m =0$ (empty space). The first one, $h(t) = 0$, corresponds to Minkowski space-time and is also the solution of the general eqs. \eqref{eq:1FE}, \eqref{eq:dh}. The second describes exponentially expanding de Sitter space with $h(t) = h_0$.


\begin{figure}[t]
	\includegraphics[width=0.7 \textwidth]{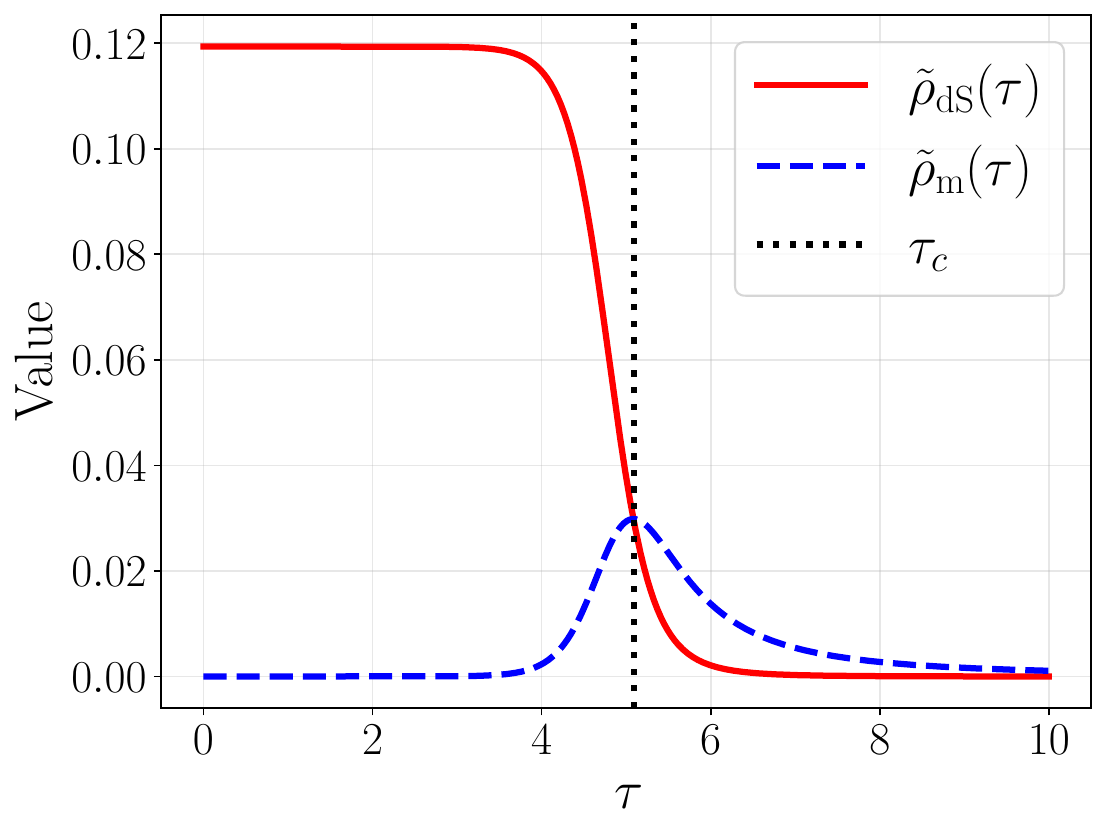}
\caption{Early cosmological history based on the solutions to \eqref{eq:1FE1} for the initial condition $h(0) = h_0(1 - 10^{-9})$.  The rescaled dimensionless time  $\tau = h_0 t$ and the dimensionless Hubble parameter $\tilde h = h/h_0$ are used here.  The quantities plotted are : the rescaled TV energy density $\tilde\rho_{\rm dS}= \frac{3}{8\pi} \tilde h^4$,  and the rescaled matter energy density $\tilde\rho_m = \frac{3}{8\pi}(\tilde h^2 - \tilde h^4)$. The curves cross at $\tau_c = h_0 t_c$ when matter is in equilibrium with TV and the Hot Big Bang begins. \la{fig:history}}
\end{figure}

The  Fig.1 shows the main features of the TVM  for  early evolution of the Universe. The initial state, say at $t=0$ is a perturbed metastable state with    $h(0) = h_0 (1 - \epsilon)$   and the infinitesimally small initial energy density of matter $\rho_m(0) \simeq  \frac{3}{4\pi} h_0^2 \epsilon$,  but very high (Planck scale) TV energy density $\rho_{\rm dS} \simeq  \frac{3}{8\pi} h_0^2$.  The first period of exponential expansion (inflation)  lasts till the time $t_c$  which can be defined as a moment when 
\be
\rho_{\rm dS} (t_c) = \rho_m (t_c) .
\la{eq:rho-eq}
\ee 
This time scales logaritmically with $1/\epsilon$  and can be interpreted as the end  of inflation and the begining of the standard Big Bang. As shown in  \cite{QTdS},  inflation last for at least 60 e-folds if  
$t_c \gtrsim \frac{60}{h_0}$, and hence $\epsilon < 10^{-124} $. At this stage of the evolution a massive production of regular and dark matter particles, at the expense of TV energy, creates  practically all matter content of the Universe.
\par
Because the number of polarizations is the same for TV and matter and both systems behave like radiation, this is the unique moment in the history of the Universe when matter is in equilibrium with TV at the Gibbons-Hawking temperature  \hbox{$T_c = h(t_c)/2\pi$}.  Moreover, $\rho_m (t)$  takes its maximum value for $t=t_c$,  as is shown on \Fig{fig:history}. Very soon  $\rho_{\rm dS}$ drops rapidly and can be neglected in comparison with the gravity contribution $ - \frac{3}{8\pi }h^2$  reducing the TVM Friedmann equations to  the standard Hot Big Bang theory, valid until the ``dark energy epoch'' when $\rho_{\rm dS}$ becomes important, again (see the next subsection).
\par
Using \eqref{eq:1FE1}  one can compute
\be
h(t_c) = \frac{h_0}{\sqrt 2} ,\quad \dot{h}(t_c) = - h_0^2 ,
\la{eq:maxh}
\ee
what implies the following values of the important parameters at the end of inflation:
\begin{enumerate}

	\item temperature  
\be
T_c \equiv T_{\rm dS} (t_c) = \frac{h(t_c)}{2 \pi} = \frac{h_0}{2 \pi \sqrt 2} ,
\la{eq:Tc}
\ee
\item  energy densities
\be
\rho_c \equiv \rho_m(t_c) = \rho_{\rm dS}(t_c) = \frac{3}{32\pi}h_0^2\ ,
\la{eq:maxdensity}
\ee
\item  particle density per single polarization (compare \cite{K&T}(eq. 3.52))
\be
n_c  = 0.1\,  T^3_c .
\la{eq:maxparticle}
\ee
\end{enumerate}
Note, that all quantities $h_0$, $T_c$ , $\rho_c$, $n_c$ and the effective duration of the rapid particle production  are of the  Planck scale.
\par
As already discussed  in the previous paper \cite{relax}, one may assume that the evolution of the  Universe has started at the metastable state corresponding to an empty de Sitter space with the Hubble constant  $h_0$. One may speculate that this initial condition has been created at the time $t=0$ by the tunneling process between Big Nothing, i.e. an empty Minkowski space,  and  Inflating Universe with the Hubble constant  $h_0$, both having null total energy density. Because presence of tunneling implies that neither Big Nothing nor Inflating Universe are exact eigenstates of the total Universe`s  Hamiltonian, intrinsic quantum fluctuations  $\epsilon \sim 1/N$ initiate TV decay accompanied by particle production. The parameter $N$  might be related to one of the large numbers emerging in cosmology like $10^{80} , 10^{104} , 10^{120}$ \cite{relax} .
\par
One should notice that the early evolution of the Universe given by \eqref{eq:1FE1} depends (beside the initial perturbation $\epsilon$) only on the form of TV energy density \eqref{eq:rhodS1} which, by dimension analysis, is universal as long as all masses of elementary particles are much lower than the Planck scale.  In this case the only  parameter of TV model, relevant at the Planck scale, is the effective number of degrees of  freedom of the theory (``polarizations''). 

\subsection{Late evolution of the Universe}
\la{sec:lateuniverse} 

It follows from the TVM equations \eqref{eq:1FE1}, \eqref{eq:dh} and is illustrated by Fig.1,  that the Hubble parameter $h(t)$ rapidly decreases in the vicinity of the time $t_c$. For times $t$ such that  $t - t_c$ is essentialy longer than the Planck scale,  $h(t) << 1$, and  the TV energy density $\rho_{\rm dS}$ can be neglected in comparison with the gravitational energy density ($- \frac{3}{8\pi }h^2$), what allows to use the standard Big Bang Theory. 
\par
Much later, when $T_{\rm dS}(t)$ is much smaller than the minimal, nonzero mass of elementary particles, the TV energy density is again relevant. Then, TV can be modelled by the extremally cold gas of elementary particles at  the particular density  for each species. This density of a quantum gas is determined by the special choice of the chemical potential - zero in the nonrelativistic formulation, or equal to the mass in the relativistic one.  Physically, it corresponds to a single particle  per  volume  $\lambda_{\rm th}^3( T_{\rm dS})$ defined by  the thermal de Broglie wavelength. 
\par
To estimate the TV energy density for the late Universe, one can use a leading term for the ideal, quantum, nonrelativistic,  spinless Bose gas  (expressed in Planck units)
\be
\rho_j  = 0.16\, m_j^{5/2} T_{\rm dS}^{3/2} .
\la{eq:mass}
\ee 
Notice, that for massless particles energy density scales like $ T_{\rm dS}^4$ and hence can be neglected, in this regime.
\par
The TV energy density is now a sum of the terms \eqref{eq:mass} for each massive particle polarization, and  using \eqref{eq:dh} with $w_m = 0$ one obtains  simplified Friedmann equations valid for the  late Universe
\be 
\rho_m = \frac{3}{8\pi}  h^2 \left[1 - \sqrt{\frac{h_{\infty}}{h}}\right], \quad  \dot h = -  4\pi \rho_m .
\la{eq:3Friednew}
\ee
with a constant
\be
h_{\infty} = 10^{-4}\left[\sum_j g_j m_j^{5/2} \right]^2 .
\la{eq:hinfty}
\ee
Here $g_j$ is a number of polarizations for the $j$-th particle  (again with a suitable correction for fermions). The eq.\eqref{eq:3Friednew} implies that $h(t)$ decreases asymptotically to the value $h_{\infty}$ and the Universe approches an empty de Sitter space-time with a very low Hubble constant. This is the third fixed point of TVM equations, which in contrast to the initial, inflating de Sitter space is stable. One can notice that  the energy density of TV (``dark energy'' or ``cosmological costant'') $\rho_{\rm dS} \simeq \frac{3}{8\pi}\sqrt{h_{\infty}} h^{3/2}$ decreases with time.
\par
Similarly to the case of  the early Universe (see \eqref{eq:1FE1}), its late evolution depends only of the scaling of TV energy density, which can be written as
\be
\rho_{\rm dS} = M^{5/2} T_{\rm dS}^{3/2} ,
\la{eq:rhodS2}
\ee
where $M$ is a mass parameter characterizing particle mass spectrum. 
\par
Using \eqref{eq:3Friednew} one can compute the value of  ``acceleration''  $\ddot{a}(t)$ for the late Universe
\be
 \frac{\ddot{a}}{a} =\dot{h} + h^2 = \frac{h^2}{2}\left( \sqrt{\frac{9 h_{\infty}}{h}} - 1\right) .
\la{eq:accel1}
\ee 
It is clear that when decreasing $h(t)$ crosses at the time $t_{\rm ac}$ the value
\be
h(t_{\rm ac}) =   9 h_{\infty},
\la{eq:accel2}
\ee
$\ddot{a}$ changes its sign and the  "accelerating expansion"  begins. 
\par
The predictions of eq.\eqref{eq:3Friednew} can be compared with  observational data introducing, instead of the cosmological time $t$, the cosmological redshift $z$, 
\be
1 + z  = \frac{1}{a(t)} .
\la{eq:red}
\ee
This change of variables transforms the differential equation \eqref{eq:3Friednew} into the equation for $h[z]\equiv h(t)$
\be
  \frac{dh[z]}{dz} = -\frac{3}{2(1+z)} \left( h[z] -\sqrt{h_{\infty}}\sqrt{h[z]}\right),
\la{eq:Friedz}
\ee
which can be solved by separation of variables getting
\be
 h[z] =  h[0]\left[\Gamma +\left(1 -\Gamma\right) (1 +z)^{3/4} \right]^2 ,
\la{eq:hzsol}
\ee
with 
\be
\Gamma = \sqrt{\frac{h_{\infty}}{h[0]}} < 1 .
\la{eq:Gamma}
\ee
It is convenient to introduce a new normalized function $f(z)$  defined by
\be
f(z) \equiv\frac{h [z]}{h[0](1 + z)},
\la{eq:function}
\ee
which in the case of accelerated expansion possesses a local minimum  at  $z_{\rm ac}$ corresponding to the moment $t_{\rm ac}$  when acceleration starts.
\par
Combining \eqref{eq:hzsol} with \eqref{eq:function} one obtains the following function  $f(z)$ characterizing TVM
\be
f(z)=\left[\Gamma (1+z)^{-1/2} + (1- \Gamma) (1 + z)^{1/4}\right]^2 .
\la{eq:f_eq}
\ee
One can easily check,  that for  $1 > \Gamma > 1/3$ the function $f(z)$ given by \eqref{eq:f_eq} possesses a minimum at 
\be
z_{\rm ac} = \left(\frac{2\Gamma}{1 - \Gamma}\right)^{4/3} - 1 .
\la{eq:fmin}
\ee
For the comparison, consider  the standard flat $\Lambda$CDM model for the late Universe, dominated by nonrelativistic matter and cosmological constant, and described by the Hubble parameter of the form \cite{K&T}
\be
h[z] = h[0] \sqrt{(1- \Omega_{\Lambda}) (1 + z)^3 + \Omega_{\Lambda}} ,
\la{eq:h_LCDM}
\ee
where $ \Omega_{\Lambda}$ is the energy density of cosmological constant relative to the critical density. The eq. \eqref{eq:h_LCDM} defines the function 
\be
f_b(z)=  \sqrt{(1- \Omega_{\Lambda}) (1 + z) + \Omega_{\Lambda}(1 + z)^{-2}} .
\la{eq:f_LCDM}
\ee
Putting $\Gamma = 0$ in \eqref{eq:f_eq}, or $ \Omega_{\Lambda}=0$ in \eqref{eq:f_LCDM})  one obtains the third case of the function $f(z)$ characterizing (nonrelativistic) matter dominated Universe
\be
f_c(z)= \sqrt{1+ z}.
\la{eq:f_matter}
\ee
Fig.2. shows  plots of the curves defined above :  $f(z)$ with $\Gamma = 0.40 , 0.45 , 0.50$ \eqref{eq:f_matter},  $f_b(z)$  with $ \Omega_{\Lambda} =0.7$ \eqref{eq:h_LCDM}, and  $f_c(z)$ \eqref{eq:f_eq}.
Visual comparison of the Fig.2 with the numerous published data (see e.g. a compilation of the earlier results in \cite{Comp} fig.1,  and the recent data of  \cite{DESI}) shows that within, still substantial, error bars the curves  corresponding to TVM and $\Lambda$CDM fit well. On the other hand the curve for  $f_c(z)$ deviates strongly from the data, and does not possess any local minimum expected due to the well established accelerated expansion \cite{Riess}, \cite{Perlmutter}. The advanced statistical analysis of  DESI results (see e.g. \cite{DESI}) favors  a  ``dark energy'' weakening over time, what agrees with the predictions of TVM.  However, error bars in the raw data remain high and the determination of  $h[0]$ is plagued by the ``Hubble tension''  of about $10 \%$  suggesting either systematic errors or necessary modification of the $\Lambda$CDM model. The later possibility has been discussed in the recent paper of the author \cite{TVHT} where the $\Lambda$CDM formula  \eqref{eq:h_LCDM} was replaced  by its TVM version \eqref{eq:hzsol} explaining the origin of the Hubble tension.

\begin{figure}[t]
	\includegraphics[width=0.7 \textwidth]{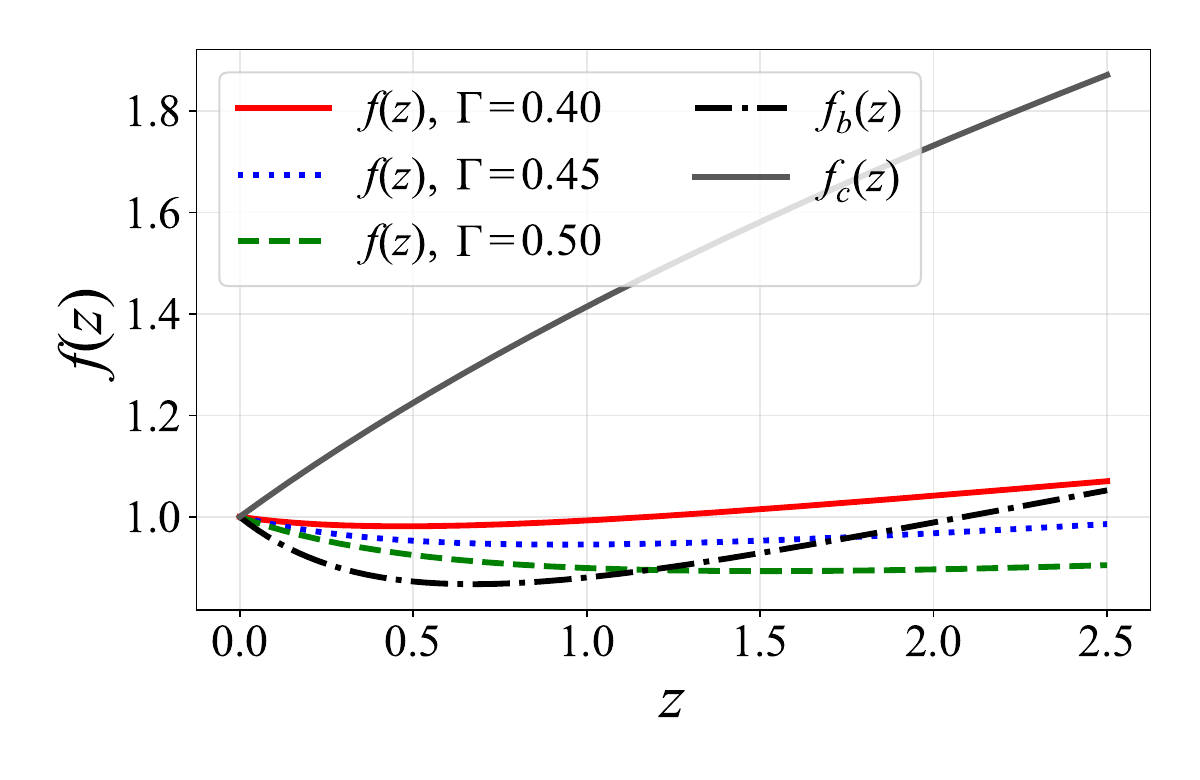}
\caption{Plots of the $f(z)$ curves   given by: a)  \eqref{eq:f_eq} with $\Gamma = 0.40 , 0.45 , 0.50$; b), \eqref{eq:h_LCDM} with $ \Omega_{\Lambda} =0.7$; c)  \eqref{eq:f_matter}}
\end{figure}

An interesting prediction of TVM applied to the late Universe is an upper bound on elementary particle mass spectrum $\bar m$ which follows from the formula \eqref{eq:hinfty}. It is clear that the value of $h_{\infty}$ is dominated by the highest mass $\bar{m}\simeq M$, (see \eqref{eq:rhodS2}), leading to the estimate
\be
\bar m \simeq  10 (h_{\infty})^{1/5}.
\la{eq:f_maxmass}
\ee

Due to \eqref{eq:Gamma} with $\Gamma \simeq 0.4$ , $h^{-1}_{\infty}$ and $h[0]^{-1}$, are both  of the order of Universe age $\sim 10^{19}s $.  Remembering that  all expressions are written in Planck units, one obtains the following estimation for the maximal mass of elementary particles forming a cold gas of the late TV 
\be
\bar m \simeq  10^{-12} M_{\rm Pl} \sim 10^7 GeV ,
\la{eq:f_maxmass1}
\ee
where $M_{\rm Pl}\simeq  1.2 \cdot 10^{19} GeV$  is the Planck mass. The Gibbons-Hawking temperature  for the late Universe, $T_{\infty} =  h_{\infty}/2\pi \sim 10^{-35}  eV$, is indeed  much below any mass estimations for massive neutrinos or any other particle mass expected in a dark matter sector, justifying the approximations leading to \eqref{eq:3Friednew}.

\section{Baryogenesis and dark matter in TVM}
\la{sec:TVMconsequences}

TVM applied to the early Universe differs very much from the existing theories because inflation and rapid transition to hot Big Bang regime, accompanied by violent particle production, happen on the Planck time scale with Planck scale values of energy densities. Therefore, the underlying fundamental theory  should be  an effective quantum theory with the cut-off energy scale $M_{*}$ much larger than the Planck mass $ M_{\rm Pl}$. Expected ``anomalous quantum gravity effects"of the order of $M_ {*}^{-2}$ should describe new physics relevant for cosmology. In the following two examples of such  effects are discussed within TVM : gravitational baryogenesis and some limits on masses for dark matter particles.

\subsection{Gravitational Baryogenesis}
\la{sec:gravitation}
Most of the proposed theoretical mechanisms for baryogenesis are based on Sakharov conditions \cite{Sakharov} invoking violation of baryon $B$ or lepton number $L$, violation of $C$ and $CP$ symmetries
and out-of-equilibrium state.   There is one qualitatively different theoretical proposal, due to Cohen and Kaplan \cite{CK-2}, who assumed that even though $CPT$ is an exact microphysical symmetry, it is spontaneously broken by the {\it time-dependent} cosmic background.  This macroscopic breaking of $CPT$ allows baryogenesis to proceed without $CP$ violation and while $B$-violating interactions are acting under temporal thermodynamic equilibrium conditions.
\par
An elegant implementation of Cohen and Kaplan idea, proposed by Davoudias et. al. \cite{gravitational} employs the effective  anomalous action 
\be
\mathcal{S}_{\rm int} = \frac{1} {M_{*}^2}\int d^4 x \sqrt{|g|}\left(\partial_{\mu}R\right) j^{\mu}_B .
\la{eq:action}
\ee
where $j^{\mu}_B$ is the baryonic  current and $R$ is the Ricci scalar.  This proposal is minimal in the sense that no other fields  are introduced and  gravity is known to violate $B$, along with any other global symmetry; see, e.g., \cite{Kallosh} and references therein.
\par
For the FLRW  metric this action induces an effective  time-dependent quantum Hamiltonian for baryons  in a finite comoving volume 
\be
\hat{H}_B (t) = \frac{1} {M_{*}^2}\dot{R}(t) \left[\hat{N}^{(+)}_B - \hat{N}^{(-)}_B\right],
\la{eq:hambar}
\ee
where $\hat{N}^{(\pm)}_B$ is the baryon or antibaryon number operator, respectively. The Ricci scalar $R$ given by
\be
R(t) = 6 \left[ \dot h (t) + 2 h^2(t) \right] ,
\la{eq:ricci}
\ee
plays the role of the time-dependent cosmic background proposed to drive baryogenesis in  \cite{CK-2}.  For the early Universe, described by \eqref{eq:1FE1}, $R(t)$ takes a simple form
\be 
R(t) = 12\frac{h^4(t)}{h_0^2}= 32\pi \rho_{\rm dS}(t) .
\la{eq:ricci1}
\ee
Fig.1 shows that the time derivative $\dot{R}(t)  \propto \dot{\rho}_{\rm dS}(t)$ is essentially zero outside the Planck scale time interval around the characteristic time $t_c$, and takes there huge,  Planck scale, values.
In this short time interval practically all matter is created and thermalizes to the temperature of TV  which can be treated as almost constant and approximately equal to $T_c$ \eqref{eq:Tc}. Therefore, the Hamiltonian \eqref{eq:hambar}, taken at $t=t_c$, produces an effect equivalent to the presence of  chemical potentials  $\pm \mu_c$ for baryons and antibaryons, respectively. The value of $\mu_c$ can be computed using \eqref{eq:hambar}, \eqref{eq:ricci1},\eqref{eq:rhodS2}, and \eqref{eq:maxh},
\be
\mu_c \equiv - \frac{\dot R(t_c)}{M_\ast^2}=  \frac{24}{\sqrt{2}}\frac{ h_{0}^3}{M_\ast^2} . 
\la{eq:muR}
\ee
One can estimate  the cut-off parameter $M_\ast $ using \eqref{eq:muR}, and  the present  photon to baryon ratio. Denoting by $n^{(\pm)}_c$ the density of baryons or antibaryons (quarks and antiquarks) at time $t_c$, one can use the following estimation for the relative excess of baryons over antibaryons caused by the difference of chemical potentials $2\mu_c$
\be
\frac{n^{(+)}_c- n^{(-)}_c}{n^{(+)}_c+ n^{(-)}_c} = \frac{2\mu_c}{T_c} = \frac{12}{\pi} \left(\frac{h_0}{M_\ast}\right)^2 .
\la{eq:b_antib}
\ee
Eq.\eqref{eq:b_antib} is valid under the condition $\mu_c << T_c$, which is satisfied when $M_\ast >> M_{\rm Pl}$ (see \eqref{eq:muR}, \eqref{eq:Tc}). Because of subsequent rapid decrease of TV temperature the equilibrium mechanism of producing the excess density of baryons ceases to work freezing the total excess of baryon number in the given comoving space volume. This approximative  baryon number conservation leads to the following relation  
\be
n^{(+)}_c- n^{(-)}_c = 3n_A a(t_c)^{-3} = 3n_A\left( \frac{T_c}{T_{\rm CMB}}\right)^{3},
\la{eq:b_antib1}
\ee
where $ n_A  \simeq 1 ~\hbox{m}^{-3} $ is the present density of  nucleons ($3n_A$ quarks)  in the Universe, while  $a(t_c)^{-3}$ accounts for the volume expansion between the begining of  Big Bang at $t_c$ and present time. The final estimation in \eqref{eq:b_antib1} follows from the relation $T \propto a^{-1}$, valid for radiation temperature, and  with $T_{\rm CMB}$ denoting the present temperature of CMB. In the next step we use the formula  \eqref{eq:maxparticle} for the particle density of radiation  to obtain
\be
n^{(+)}_c + n^{(-)}_c = 0.1 g_b T_c^3, \quad   n_{\gamma}= 0.2 T_{\rm CMB}^3, 
\la{eq:raddensity}
\ee
where $g_b$ is the number of barionic polarizations and $n_{\gamma}$ is the present density of CMB photons.  Inserting \eqref{eq:raddensity} into \eqref{eq:b_antib1} and then into  \eqref{eq:b_antib} one obtains the final estimation
\be
M_\ast = \sqrt{\frac{20 g_b}{\pi} \frac{n_{\gamma}}{n_A}} h_0 \simeq 10^6 M_{\rm Pl} ,
\la{eq:cutoffM}
\ee
where  $g_b \simeq 10^2$ and  $ n_{\gamma}/n_A \simeq 10^{10}$ is the present photon to baryon ratio, justifying the initial assumption  $M_\ast >> M_{\rm Pl}$ .
\par


\subsection{Dark Matter}
\la{sec:darkmatter}

There exists a strong evidence that 85\% of matter consists of massive particles which are not described by the Standard Model (SM). Their interaction with SM particles seems to be only gravitational, what includes also quantum gravity effects similar to those which may be responsible for baryogenesis. In particular, if  dark matter (DM) consists of singlets with respect to any gauge symmetry they are unprotected and decay into SM particles due to anomalous interactions. This mechanism has been applied in \cite{darkmatter} to  derive upper bounds on masses for ``stable'' DM particles with life-times longer than the age of Universe. The authors of   \cite{darkmatter} estimated life-times using the anomalous actions proportional to $1/M_{\rm Pl}$ which produce decay rates $\Gamma_{\phi}\propto m_{\phi}^3/M_{\rm Pl}^2$ or  $\Gamma_{\psi}\propto m_{\psi}^5/M_{\rm Pl}^2$  for a scalar boson ($\phi$)  or a spin-1/2 neutral fermion ($\psi$), respectively. Because TVM suggests that the cut-off energy scale is rather $M_\ast \simeq 10^6  M_{\rm Pl}$ one can rescale the results of \cite{darkmatter} to obtain the higher upper bounds on  masses for ``stable''  scalars  and  fermions
\be 
m_{\phi} < 10^2 GeV, \quad  m_{\psi} < 10^3 GeV .
\la{eq:dmatterlife}
\ee
On the other hand, due to the estimation \eqref{eq:f_maxmass1}, DM sector must contain at least one heavy particle with the mass $\bar{m}\simeq 10^7  GeV$. To suppress the decay of heavier DM particles into SM particles one can assume the existence of at least one scalar boson which can mediate, through Yukawa interaction, the following  processes dominating over decays into SM sector: 

i) decay of a heavier fermion into a lighter one and a boson,

ii) annihilation of a heavier fermion pair into a  (virtual) boson and subsequent creation of  a lighter pair,

iii) decay of a boson into a pair of light fermions.

The lightest fermions  can still  annihilate into a  (virtual) boson which by the anomalous interaction can decay into a pair of photons allowing the indirect observation of DM \cite{DMobs}.
\par
Taking into account the limitations imposed by TVM one can consider as a plausible proposal the following structure of DM sector.  It may be treated as  a``singlet counterpart" of SM sector consisting of about $10^2$  ``dark baryons", i.e.  spin-1/2 neutral fermions, and a single neutral scalar -``dark higgs" mediating  Yukawa interaction between dark baryons.  If, additionally, the dark higgs mass and the Yukawa coupling produce a force with a strength comparable to the weak nuclear force one arrives at  the well known model of Weakly Interacting Massive Particles (WIMPs) \cite{WIMP}. An alternative scenario assumes self-interaction  mediated by Yukawa coupling with  large scattering cross-sections leading to the Self-Interacting Dark Matter (SIDM) model \cite{SIDM}
 

\section{Concluding remarks}
\la{sec:conclusions}

The following theoretical predictions indicate, still poorly understood, link between general relativity, quantum field theory and quantum thermodynamics:

a) a black-body \emph{Hawking radiation} released outside a black hole's event horizon,

b) the \emph{Unruh effect}  for an uniformly accelerating observer who will perceive an empty space as a thermal bath,

c) a quantum vacuum in expanding de Sitter space, characterized entirely by the \emph{Gibbons-Hawking temperature}, seen as a thermal equilibrium state by an observer resting in the cosmic comoving reference frame.
\par
As discussed in this paper, the last effect combined with the standard  FLRW model of the Universe yields a feasible cosmology dependent on few parameters only. This model describes inflation period, the transition to Hot Big Bang scenario, and accelerated expansion for the late Universe without introducing inflaton field(s), reheating mechanism and cosmological constant. The varying by a huge factor                 $[h_0/h_{\infty}]^2 \sim 10^{120}$  TV  energy density solves, in a natural way, the \emph{cosmological constant problem}. 
\par
It remains to develop a theory of  primordial cosmological perturbations  based on TVM,  which refers to thermal fluctuations of expanding vacuum, instead of ill-defined zero point quantum fluctuations of inflaton field (see \cite{noise} for the first attempt in this direction). An ultimate challenge is a search for fundamental theory which can provide  a background for TVM. As already noticed in \cite{relax} the TVM Friedmann equations \eqref{eq:1FE1} are similar to those describing irreversible superfluorescence phenomena in quantum optics. Because Friedmann equations are highly symmetric version of Einstein equations for General Relativity (GR), it may suggest that the classical GR equations  are an  effective, semi-classical and Markovian approximation to certain irreversible dynamics of an unknown quantum open system. This idea differs from the standard interpretation of GR  as classical  limit of  Hamiltonian quantum gravity. The fact that the generic model of superfluorescence consists of many ``qubits''  interacting with bosonic field may help in establishing connections of TVM to  other approaches in the field of quantum gravity.


{\bf Acknowledgements}:  The author thanks Gabriela Barenboim and Alejandro Jenkins for valuable discussions, and Borhan Ahmadi for making figures .


\end{document}